\documentstyle[12pt]{article}
\def\be{\begin{equation}}
\def\ee{\end{equation}}

\begin{document}
\setcounter{page}{0}
\renewcommand{\thefootnote}{\fnsymbol{footnote}}
\thispagestyle{empty}
\begin{flushright}
TIFR-TH/96-44\\
BU-TH/96-3\\
{\bf hep-ph/9608352} \\
\end{flushright}
\vskip 45pt
\begin{center}
{\Large \bf \boldmath Bounds from $t \bar t$ production on 
$R$-parity violating models of supersymmetry}

\vspace{11mm}
\bf
 Dilip Kumar Ghosh${}^{(1)}$\footnote{dghosh@theory.tifr.res.in},
 Sreerup Raychaudhuri${}^{(2)}$\footnote{sreerup@theory.tifr.res.in}
 and K. Sridhar${}^{(2)}$\footnote{sridhar@theory.tifr.res.in} \\

\vspace{13pt}
 {\sl (1) Department of Physics, University of Bombay,
   Vidyanagari, Santa Cruz (East), Bombay 400 098, India.\\
       (2) Theory Group, Tata Institute of Fundamental Research,
   Homi Bhabha Road, Bombay 400 005, India.}

\vspace{50pt}
{\bf ABSTRACT}
\end{center}

\begin{quotation}
We study $t \bar t$ production in $R$-parity violating supersymmetry.
The annihilation channel $q \bar q \rightarrow t \bar t$ gets new
contributions from $t$-channel exchange of squarks or sleptons. 
With the data from Tevatron on $t \bar t$ production, we find that the
squark- or slepton-exchange processes constrain the $B$-violating 
$\lambda''$ couplings or the $L$-violating $\lambda'$ couplings, 
respectively. Our bounds are already comparable to the few existing 
constraints on third-generation $R$-parity violating couplings, and
will improve when more precise measurements of the $t \bar t$ production
cross-section become available. We also discuss the effects of these
couplings for top production at the LHC.
\end{quotation}

\vspace{1.25in}

\vfill
\newpage
\setcounter{footnote}{0}
\renewcommand{\thefootnote}{\arabic{footnote}}

\setcounter{page}{1}
\pagestyle{plain}
\advance \parskip by 10pt

Supersymmetry is considered to be one of the most promising candidates 
for physics beyond the Standard Model (SM) and, consequently, a 
significant amount of effort has been devoted to looking for signals of 
supersymmetry in experiments. The minimal supersymmetric extension of 
the Standard Model (MSSM) \cite{mssm} contains, in addition to the usual 
particles of the Standard Model, their superpartners and two Higgs 
doublets. While the gauge structure of the MSSM essentially replicates 
that of the Standard Model, the Yukawa sector of the MSSM is somewhat 
more complicated. In addition to the usual Yukawa couplings of the 
fermions to the Higgs (responsible for the fermion masses), other 
interactions involving squarks or sleptons are possible.

\vskip10pt
The relevant part of the superpotential containing the Yukawa interactions
involving squarks or sleptons in the MSSM is given in terms of the
chiral superfields by
\be
    {\cal W}= \lambda_{ijk} \epsilon_{\alpha \beta} L_i^{\alpha} 
    L_j^{\beta} E^c_k +  \lambda'_{ijk} \epsilon_{\alpha \beta}
    L_i^{\alpha} Q_j^{\beta} D^c_k + \epsilon_{\alpha\beta} \mu_i 
    L_i^{\alpha} H_2^{\beta} + 
    \lambda''_{ijk} \epsilon^{abd} U^c_{ia} D^c_{jb} D^c_{kd}
      \label{e1}
\ee
where the $L_i$ and $Q_i $ are $SU(2)$-doublet lepton and quark superfields
and the $E_i, U_i, D_i$ are singlet superfields, $H_2$ is a Higgs 
superfield, $c$ denotes conjugation, $i,\ j,\ k$ are generation indices, 
$\alpha, \beta$ are $SU(2)$ indices and $a,\ b,\ d$ are $SU(3)$ indices. 
The couplings $\lambda$, $\lambda'$ and $\mu_i$ violate lepton ($L$) number,
whereas the $\lambda''$ coupling violates baryon ($B$) number. The $B$- 
and $L$-violating couplings cannot be present simultaneously, because 
that would lead to very rapid proton decay. The term $\epsilon_{\alpha\beta} 
\mu_i L_i^{\alpha} H_2^{\beta} $ is not usually included, because it can
be rotated away from the superpotential by a redefinition of the Higgs $H_1$
and the leptonic $L_i$ superfields \cite{hall} \footnote{We remark that
this redefinition does not leave the full Lagrangian (i.e. including
the soft supersymmetry-breaking terms) invariant and can be achieved only
at the expense of generating two additional complex parameters in the
soft supersymmetry-breaking sector of the Lagrangian \cite{anjan}. This
complication is of no consequence to our analysis.}. We will not consider
this term any further. It is possible to forbid the existence of all the 
interactions in Eq.~(\ref{e1}) by imposing a discrete symmetry -- $R$-parity. 
This discrete symmetry may be represented by $R=(-1)^{(3B+L+2S)}$,
where $S$ is the spin of the particle, so that the usual particles of
the SM have $R=1$, while their superpartners have $R=-1$.

\vskip 10pt
The requirement that the MSSM Lagrangian be invariant under $R$-parity
is sufficient to exclude each of the interactions in Eq.~(\ref{e1}).
However, $R$-parity conservation is too strong a requirement to ensure 
proton stability \cite{rparv}~-- the latter can simply 
be ensured by assuming that either the $L$-violating or the 
$B$-violating couplings in Eq.~(\ref{e1}) are present, but not both.
Relaxing the requirement of $R$-parity conservation has important
implications for supersymmetric particle searches at colliders~:
a superparticle can decay into standard particles {\it via} the 
$R$-parity violating couplings in Eq.~(\ref{e1}), and, hence, the
lightest supersymmetric particle will no longer be stable or escape
detection. Thus bounds on MSSM parameters derived from missing energy
and momentum signals are no longer valid in $R$-parity violating scenarios.

\vskip10pt
In this letter, we will not be concerned with the $L$-violating
$\lambda$ couplings, since we will be interested in the effects
of $R$-parity violation at hadronic colliders.
Consequently, we focus on $\lambda'$ and $\lambda''$ couplings
and we begin by presenting a quick review of the existing bounds
on the $\lambda'$ and $\lambda''$ couplings. Direct searches for
the squarks coupling to quarks and leptons via the $L$-violating 
interaction induced by the $\lambda'$ coupling,
by the CDF \cite{cdf} and the D0 experiments \cite{d0} at the Tevatron 
yield bounds \cite{dp} on the first generation squark which are of the
order of 100~GeV (almost irrespective of the size of the Yukawa couplings).
At HERA, the H1 collaboration has excluded \cite{h1} first-generation 
squarks upto masses of 240~GeV for a $\lambda' \sim \sqrt{4 \pi 
\alpha_{\rm em}}$. There are also reasonably stringent indirect
limits on first- and second-generation $\lambda'$ couplings. For
instance, the upper limit on the $\nu_e$ majorana mass \cite{dimo} 
or the non-observation of neutrinoless double beta decay \cite{rm}
give $\lambda' < 10^{-2}$ for a squark mass of 100~GeV. Bounds of
the order of $10^{-2} - 10^{-1}$ on $\lambda'$ from a variety of 
low-energy observables have been obtained \cite{bgh}. In contrast, 
there are weaker bounds on the third-generation couplings and the 
most stringent limits are of the order of 0.5 which are obtained by 
considering radiative corrections induced by $\lambda'$ couplings
to the partial widths of the $Z$ \cite{bes}.

\vskip10pt
For $\lambda''$ couplings involving the first-generation, the non-observation
of $n \bar n$-oscillations \cite{nnbar} and double nucleon decays into
kaons of identical strangeness \cite{kaon} place very strong constraints
on the $\lambda''$ couplings. Data on neutral meson mixing or that on
$CP$-violation in the $K$-sector can also be used to put stringent
bounds on certain products of the $\lambda''$ couplings. Again, as for
$\lambda'$, the $\lambda''$ couplings involving third-generation quarks 
are not very strongly constrained~-- the best bounds \cite{bcs} are of 
the order of 1.25 for a squark mass of 100~GeV, obtained from computing 
the contribution to $R_l=\Gamma_l/\Gamma_h$ from one-loop diagrams 
involving squarks and top-quarks. Similar bounds on $R$-violating couplings 
involving third-generation quarks also follow from the assumption of 
perturbative unification \cite{unifi}. Cosmological considerations
seem to place very strong constraints~-- for example, the
requirement that GUT-scale baryogenesis does not get washed
out gives $\lambda''<< 10^{-7}$ generically \cite{cosm}, though these 
bounds are model dependent and can be evaded \cite{dr}. 

\vskip10pt
Given that the constraints on $\lambda'$ and $\lambda''$ couplings 
involving third-generation fields are not very stringent, it is 
important to study $L$- and $B$-violating processes that involve 
third-generation particles. In this letter, we study $t \bar t$ production
at the Tevatron with precisely this objective in mind. In addition,
to the usual QCD sub-processes that contribute to $t \bar t$ 
production, we have the distinctive $t$-channel slepton/squark exchange
subprocess also contributing, when the corresponding $R$-violating
coupling is non-zero. The measured values of the cross-section
from the CDF and the D0 experiments at the Tevatron allows us to
put constraints on the third-generation $R$-violating coupling as
a function of the slepton or squark mass.

\vskip10pt
Since we are only interested in the interactions generated by the
$\lambda'$ and $\lambda''$ couplings, we write this part of the 
interaction in Eq.~(\ref{e1}) in terms of the component fields. 
In four-component Dirac notation, the part of the Lagrangian 
with $\lambda'$ couplings is given by
\begin{eqnarray}
    {\cal L}_{\lambda'} =&  - \lambda'_{ijk} \lbrack
                & \tilde{\nu}_L^i \bar{d}_R^k d_L^j +
                \tilde{d}_L^j \bar{d}_R^k \nu_L^i +
                (\tilde{d}_R^k)^* (\bar{\nu}_L^i)^c d_L^j
 \nonumber \\
                &&- \tilde{e}_L^i \bar{d}_R^k u_L^j -
                \tilde{u}_L^j \bar{d}_R^k e_L^i -
                (\tilde{d}_R^k)^* (\bar{e}_L^i)^c u_L^j
                                \rbrack  + \mbox{\rm h.c.} ;
\label{lagrp1}
\end{eqnarray}
and the corresponding expression for the $\lambda''$ coupling:
\be
{\cal L}_{\lambda''}
= - \lambda''_{ijk} \left[ \tilde d_R^k (\bar u^i_L)^c d_L^j
+\tilde d_R^j (\bar d^k_L)^c u_L^i
+\tilde u_R^i (\bar d^j_L)^c d_L^k \right] + 
{\rm h.c.}.
\label{lagrp2}
\ee
Because of the existence of these couplings, we have new $q \bar q$
annihilation processes contributing to $t \bar t$ production, in addition
to the usual QCD mechanisms. These new processes involve $t$-channel
squark or slepton exchange and, as is evident from Eqs.~(\ref{lagrp1}) 
and (\ref{lagrp2}), only d-quark-initiated subprocesses contribute in 
these production channels. It is important to emphasise that proton-decay
non-observation constrains us to consider only one of either $\lambda'$
or $\lambda''$ couplings to be non-zero. In what follows, we will
present a general discussion which can be easily specialised to either of
the two cases~: this is, in fact, possible because the only difference
in the two cases is that for the $L$-violating case a slepton is exchanged
and for the $B$-violating case a squark is exchanged. In both cases, it
is only the $d \bar d$ initial state which gives a $t \bar t$ final state.
Consequently, except for an overall colour factor, there is no difference
in the expressions for the cross-sections for these two cases.

\vskip10pt
It is straightforward to compute the expression for the full cross-section
at leading order. The expressions for the cross-sections at leading
order in QCD for both the $q \bar q$- and the $gg$-initiated subprocesses
are well-known. Since the $R$-parity violating contribution 
is only via the $q \bar q$ initial state, the $gg$ part of the cross-section 
coming from QCD is unaffected by this new contribution. The $q \bar q$ 
contribution from both production mechanisms will interfere~: in fact, 
as explained above, even here only the $d \bar d$-initiated contribution 
will be affected. We write down explicitly the expressions for the 
$q \bar q$ part of the cross-section. We use the notation $\Lambda$
to denote the generic coupling, with $\Lambda$ being understood to
be $\lambda'$ or $\lambda''$, depending on the case under consideration.
\begin{eqnarray}
{d\hat \sigma_{uu} \over d \hat t} & = & T_{11}^{uu}, \nonumber  \\
{d\hat \sigma_{dd} \over d \hat t} & = & T_{11}^{dd} + T_{22}^{dd}
        + T_{12}^{dd}, 
\end{eqnarray}
where
\begin{eqnarray}
T_{11}^{uu} & = & T_{11}^{dd}
 = {16 \pi \alpha_s^2 \over 3 \hat s^4} C_{11} \lbrack 2m_t^2 \hat s
+ (\hat t -m_t^2)^2 + (\hat u -m_t^2)^2 \rbrack , \nonumber \\
T_{22}^{dd} & = & {\Lambda^4_{3i1} \over 8 \pi \hat s^2} C_{22}
{(\hat t - m_t^2)^2 \over (\hat t - \tilde m_i^2)^2} , \nonumber  \\
T_{12}^{dd} & = & -{\alpha_s \Lambda^2_{3i1} \over 2 \hat s^3
(\hat t - \tilde m_i^2)} C_{12} \lbrack m_t^2 \hat s + (\hat t -m_t^2)^2
\rbrack , 
\label{xsec}
\end{eqnarray}
and
\begin{equation}
C_{11}={1 \over 12} \Big({1 \over 12}\Big),\hskip6pt 
C_{22}={2 \over 3}\Big(1\Big),\hskip6pt 
C_{12}=-{4 \over 9} \Big({4 \over 9}\Big),\hskip6pt 
{\rm for}~ \lambda''~(\lambda') {\rm couplings}.
\end{equation}
The hadronic cross-sections are obtained by folding these sub-process
cross-sections with the parton luminosities, as follows~:
\be
\sigma_{qq} = \int dx_1 dx_2 d\hat t 
\lbrack f_{p/q}(x_1) f_{\bar p/\bar q}(x_2) + f_{p/\bar q}(x_1) 
f_{\bar p/q}(x_2) \rbrack {d\hat \sigma_{q\bar q} \over d\hat t} ,
\ee
where $f_{h/i}$ denotes the probability of finding a parton $i$
inside a hadron $h$. The gluon-gluon fusion part of the cross-section
is unaffected by the $R$-violating interactions.

\vskip10pt
Just as the presence of the $R$-violating couplings affects the
$t \bar t$ production vertex, their presence will be also felt at 
the decay vertices. As opposed to the two-body decay of the top
in the standard model, the $R$-violating channel will necessarily
involve a three-body final state when the squark or the slepton 
is heavier than the top quark ($m_t \sim 169-176$~GeV). This is 
because, for the case of $\lambda'' (\lambda')$ couplings, the top 
will decay into a virtual squark (slepton) and a $d$-type quark and the
virtual squark (slepton) will further decay into two other particles. This
three-body decay is expected to be negligibly small as compared
to the Standard Model decay, so that the branching ratio into the
Standard Model decay channel will still be close to a 100\%. This
will, however, no longer be true when the squark (slepton) mass
is smaller than the mass of the top, in which case the decay of the
top quark into a $d$-type quark and a $real$ $d$-type squark (slepton) 
becomes possible. This decay mode will compete with the usual $t 
\rightarrow bW$ decay mode of the Standard Model. The ratio of these 
two decay-widths turns out to be
\be
R  \equiv  {\Gamma (t \rightarrow \bar d_i \tilde d_{jR}(\tilde e_{jL})) 
\over \Gamma (t \rightarrow bW)} \nonumber \\
= {2 s_W^2 \Lambda^2_{3ji} \over \pi \alpha} 
C_R \sqrt{ \lambda(m_t^2, m_i^2, \tilde m_j^2) \over
\lambda(m_t^2, m_W^2, m_b^2)} {m_t^2+m_i^2-\tilde m_j^2 \over
m_t^2-2m_W^2+m_b^2+{(m_t^2-m_b^2)^2 \over m_W^2}} ,
\ee
where 
\begin{equation}
C_{R}=2 (1),\hskip6pt {\rm for}~ \lambda''~(\lambda') ~{\rm couplings},
\end{equation}
and $\lambda(x,y,z) = x^2+y^2+z^2-2xy-2yz-2zx$.
The branching ratio for the decay into the $Wb$ channel then gets 
multiplied by a factor $B=(1+R)^{-1}$, when the squark (slepton)
mass is lesser than the top mass.

\vskip10pt
With the above expressions at hand, we are now able to compute
the cross-section for $t \bar t$ production in the $R$-parity
violating scenario. Before we discuss the results of our computation,
a few remarks about the significance of higher-order corrections
are in order. The cross-sections presented above are at the lowest
order in perturbation theory. In the QCD case,
significant progress has been made in computing higher-order
corrections to heavy quark production. Not only have the 
next-to-leading order corrections been calculated a long time 
ago \cite{nason}, but the resummation of soft gluons and 
its effect on the total cross-section have been recently 
computed \cite{berger}. In principle, a reliable estimate 
of the cross-section for the $R$-parity violating case 
under consideration, can also be made only when we have 
(at least) the corrections to these processes at next-to-leading 
order. But for want of such a calculation, the best we can
do is to use the leading order QCD and the resummed QCD
cross-sections \cite{berger} in the $q \bar q$ annihilation channel 
to extract a `$K$-factor' for the annihilation channel. We
then work with the approximation that the $R$-violating cross-section
will also be affected by QCD corrections in a similar fashion so that
we can fold in our cross-sections for the $R$-violating case by the 
same $K$-factor. Clearly, more work is needed in this direction
but we do not expect that our results will be qualitatively changed by
higher order QCD corrections.

\vskip10pt
In Fig.~1, we have presented the $t \bar t$ production cross-section 
for the $B$-violating $\lambda'' \ne 0$ scenario. The case of a
non-vanishing $\lambda'$ yields very similar results and we
do not present the results for the cross-section in that case separately. 
We have plotted the $t \bar t$ production cross-section as a function 
of the $\lambda''$ coupling for different values of the mass of the right 
$d$-squark. The cross-section is computed for both the Tevatron 
($\sqrt{s}=1.8$~TeV) and for the LHC energy ($\sqrt{s}=14$~TeV).
These are shown by solid and dashed lines respectively, in Fig.~1.
For the input parton distributions, have used the CTEQ-3M distributions
\cite{cteq}. This choice is motivated by the fact that our $K$-factor 
is extracted from the higher order QCD calculations presented in 
Ref.~\cite{berger} which use this set of distribution functions.
Our parton distributions were taken from the package PDFLIB \cite{pdflib}.
The 95\% confidence-level band from the D0 experiment \cite{topcs} 
is indicated by the dotted lines in Fig.~1. This is shown for
the two cases when the squark mass (100~GeV) is lighter than the top mass 
and for the case when it is heavier (marked). For a squark mass of 
100~GeV a band is disallowed for $\lambda''$, whereas for values of
squark mass higher than $m_t$, we get an upper bound.

\vskip10pt
For the results shown for the LHC energy, we have not folded in the 
cross-sections with a $K$-factor since the expectation is that at these 
energies higher-order QCD corrections will be very small. For this 
computation, we have used the MRSD-${}'$ parton distribution functions 
\cite{mrs}. Since the effect we are studying comes from a $q \bar q$ 
annihilation process, we would naively expect that the effect at the LHC 
energy is very small since we would expect that the gluon-initiated 
processes swamp the $q \bar q$-initiated process completely. While
the gluon-initiated contribution is much larger because
of the steep gluon distributions at low-$x$, the usual annihilation
process in QCD is further suppressed because it is an $s$-channel
process. The $t$-channel subprocess in the $R$-parity violating case
does not suffer this suppression and hence the effects at the LHC
energy, while smaller than the corresponding effects at the Tevatron,
are still sizeable.

\vskip10pt
Our results for the cross-sections can be easily translated into bounds
on the values of the squark/slepton masses and the corresponding
$\lambda''$ or $\lambda'$ couplings. In Fig.~2, we have shown the 
allowed regions in the $\lambda''-m_{\tilde d_R}$ plane allowed at 95\%
confidence level by the D0 and the CDF data \cite{topcs} 
on $t \bar t$ production cross-section. The bounds resulting from the
CDF data are shown by solid lines and those from the D0 data are shown by
dashed lines. For squark masses lesser than the top mass, the bounds are 
weakened considerably because the decay channel $t \rightarrow d 
\tilde d_R$ opens up, and this tends to dilute the branching into the
Standard Model decay mode $t \rightarrow bW$.

\vskip10pt
In Fig.~3, we have shown the allowed regions in the $\lambda'-m_{\tilde e_L}$
plane. We would like to point out again at this stage that the part of
the $L$-violating interaction that the $t \bar t$ production process
selects out involves only left-handed charged sleptons and does not
involve sneutrinos. As in Fig.~2, the solid and dashed lines are the 95\% bounds
resulting from the CDF and the D0 data respectively. These bounds are 
stronger than those in Fig.~2 because of the larger colour factor 
that appears in the $t \bar t$ production through $\lambda'$ couplings. 
Simultaneously, the opening up of the decay channel $t \rightarrow d 
\tilde e_L$ for light sleptons does not affect the bounds quite as 
much as in Fig.~2 because the relevant $R$-violating branching 
ratio is suppressed by a smaller colour factor. 

\vskip10pt
In summary, we have studied the constraints coming from $t \bar t$
production at the Tevatron on $\lambda'$ and $\lambda''$ couplings
in $R$-parity violating models of supersymmetry. These direct bounds that
we obtain constrain the third-generation couplings~-- and are comparable
to the indirect bounds on third-generation couplings obtained from
electroweak precision data from LEP. We stress that our aim in this paper
is to illustrate the kind of bounds that can be obtained by considering
this process. The actual bounds can be improved when the cross-section
for the production of $t \bar t$ is determined more accurately, and the
present discrepancy between the CDF and D0 values for the top cross-section
is resolved. It is also important to consider QCD corrections to the
$R$-violating processes before we can have a true quantitative estimate
of the bounds. Nevertheless, the bounds on both couplings, and especially that
obtained for $\lambda'$, are interesting and deserve more attention. Our
results for the LHC are also rather encouraging. 

{\sl \underline{Acknowledgment:} DKG acknowledges financial support from the
University Grants Commission, Government of India.

\vfill
\newpage

\begin{figure}[htb]
\vskip 8in\relax\noindent\hskip -1in\relax{\includegraphics{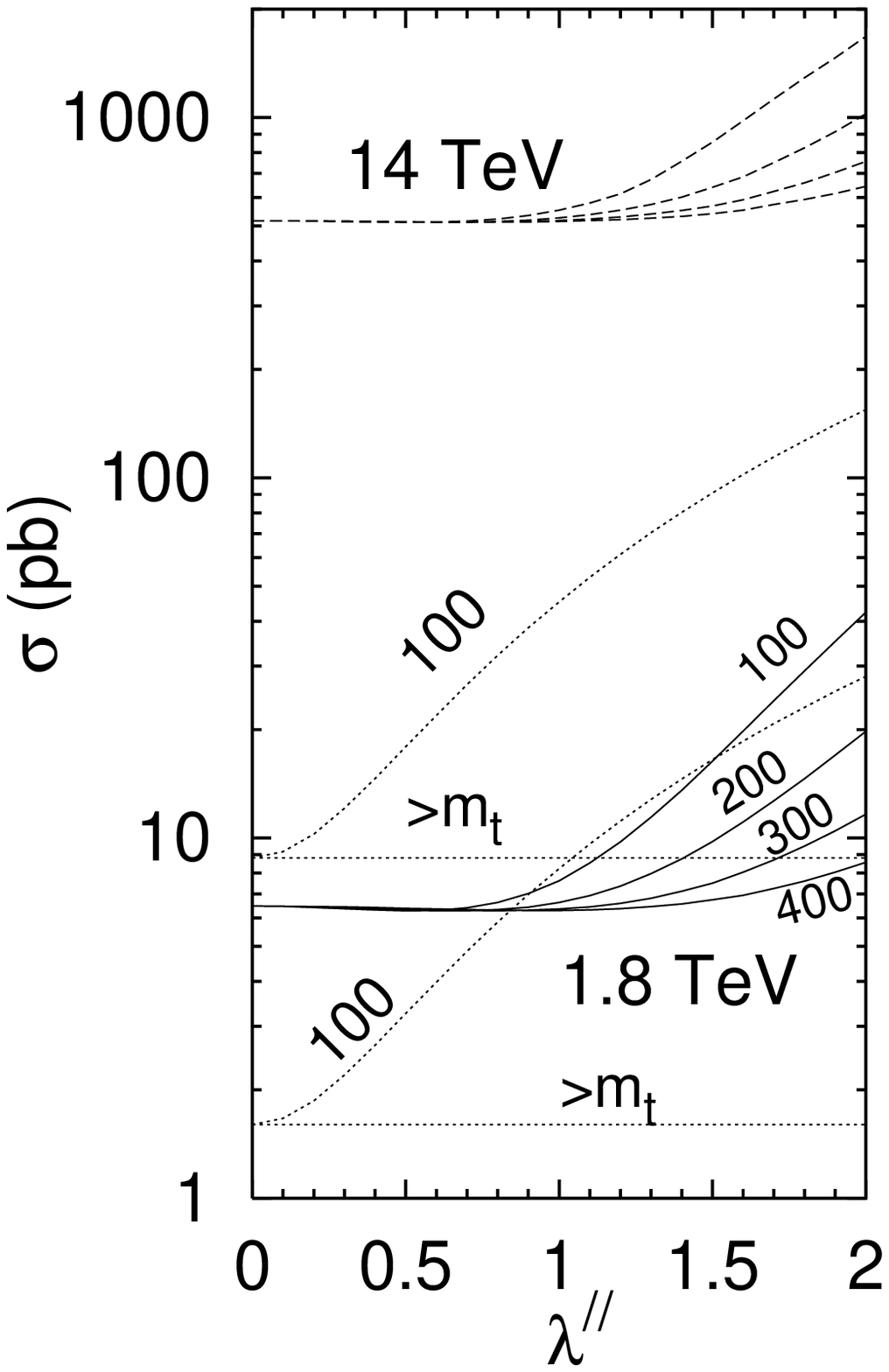}}

\vspace{-20ex}
\caption{ The $t \bar t$ production cross-section in a
        baryon-number violating scenario as a function of the 
        $\lambda''$ coupling for different values of the mass of 
        the right $d$-squark. Solid (dashed) lines show the 
        cross-section at the Tevatron (LHC), with the squark
        masses in GeV marked next to the former. Dotted lines
        show the D0 2-$\sigma$ results for the two cases when
        $m_{\tilde d_R} < m_t$ (100 GeV) and when  $m_{\tilde d_R} > m_t$ 
        (marked). }
\end{figure}

\begin{figure}[htb]
\vskip 8in\relax\noindent\hskip -1in\relax{\includegraphics{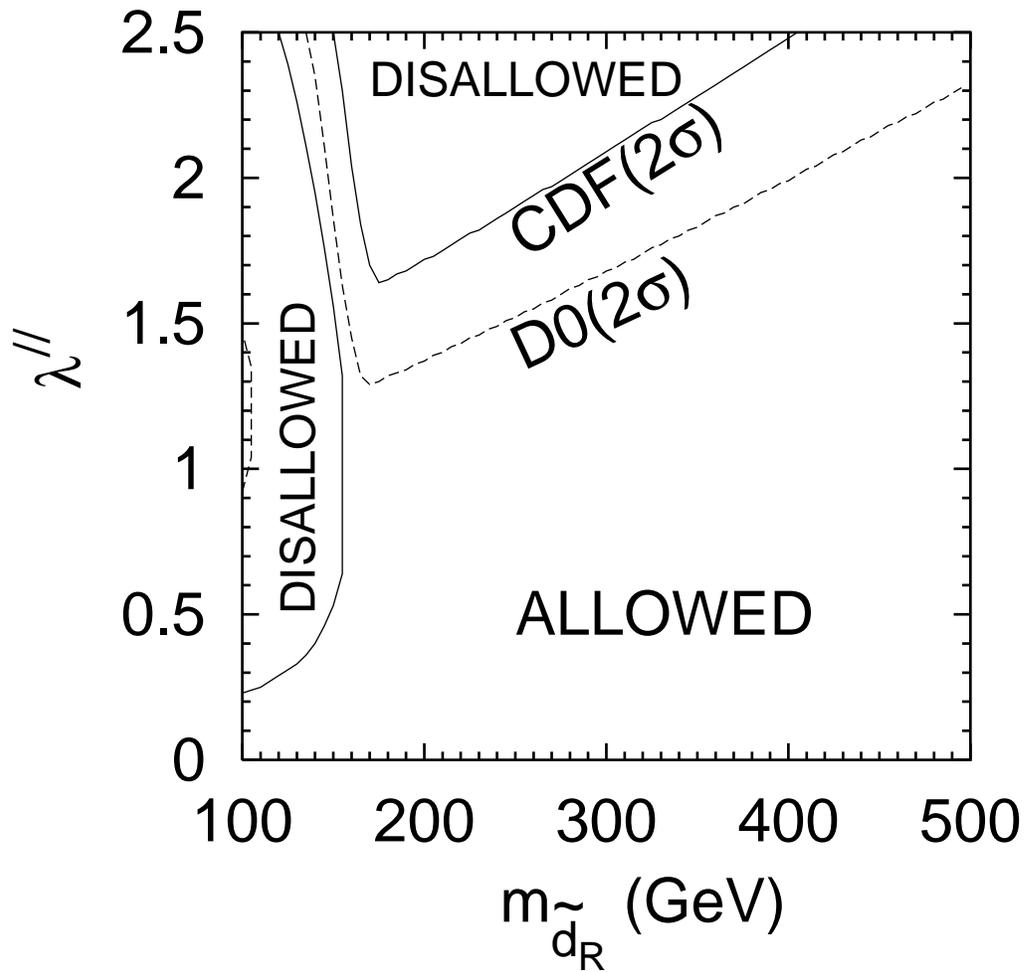}}

\vspace{-20ex}
\caption{ Allowed regions in the plane of $\lambda''$ and the mass of the
        right $d$-squark in a baryon number-violating scenario. Solid 
        (dashed) lines correspond to the 2-$\sigma$ bounds from the 
        CDF (D0) collaborations. }
\end{figure}

\begin{figure}[htb]
\vskip 8in\relax\noindent\hskip -1in\relax{\includegraphics{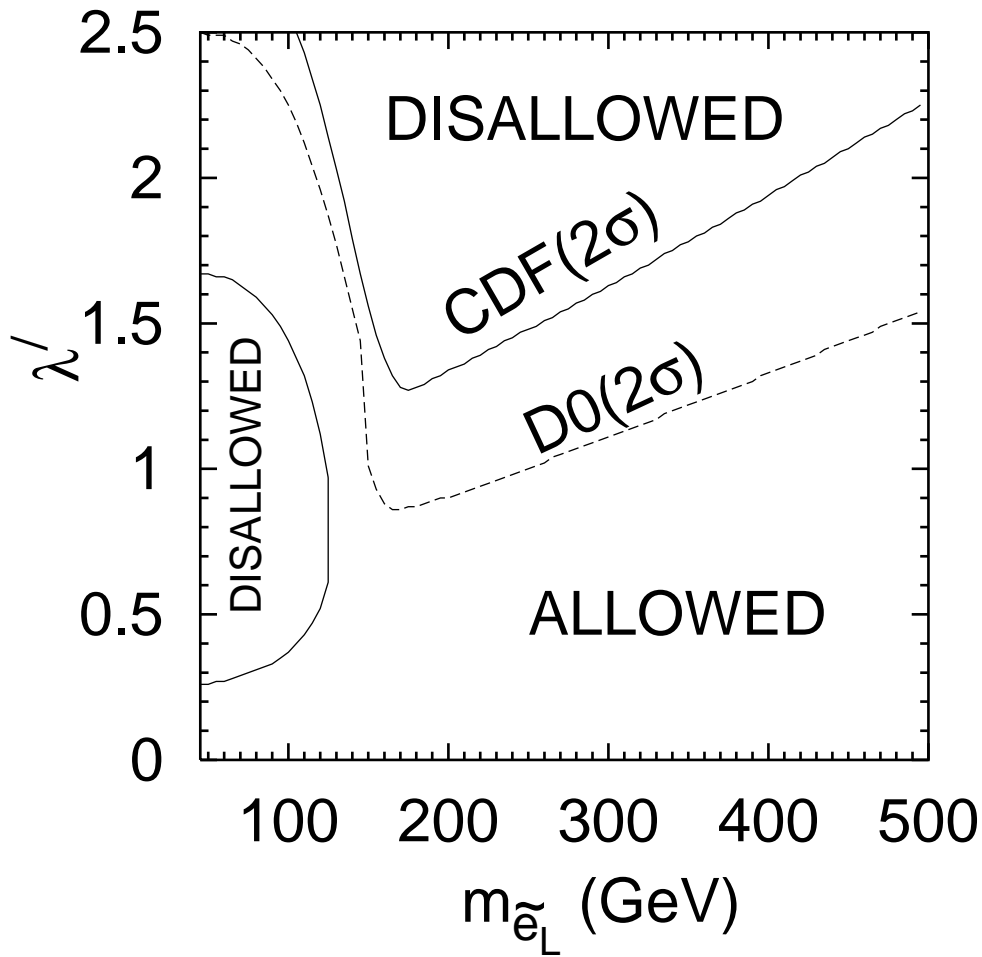}}

\vspace{-20ex}
\caption{ Allowed regions in the plane of $\lambda'$ and the mass of the
        left slepton in a lepton number-violating scenario. Solid 
        (dashed) lines correspond to the 2-$\sigma$ bounds from the 
        CDF (D0) collaborations. }
\end{figure}

\end{document}